# Multiscale simulations guided advances for all-optical phase-change waveguides

Hanyi Zhang[1], Wanting Ma[1], Wen Zhou[1,*], Xueqi Xing[1], Junying Zhang[1], Tiankuo Huang[1], Ding Xu[1], Xiaozhe Wang[1,*], Riccardo Mazzarello[2], En Ma[1], Jiang-Jing Wang[1,*], Wei Zhang[1,*]

[1]Center for Alloy Innovation and Design (CAID), State Key Laboratory for Mechanical Behavior of Materials, Xi'an Jiaotong University, Xi'an, China.
[2]Department of Physics, Sapienza University of Rome, Rome 00185, Italy

Emails: wen.zhou@xjtu.edu.cn, wangxiaozhe@xjtu.edu.cn, j.wang@xjtu.edu.cn, wzhang0@mail.xjtu.edu.cn

**Abstract**
Photonic computing using chalcogenide phase-change materials (PCMs) is under active development for energy-efficient artificial intelligence (AI) applications. A key requirement is to enable as many optically programmable levels per device as possible, while maintaining relatively low optical loss. In this work, we carry out multiscale simulations using density functional theory and finite-difference time-domain methods, proposing a "the shorter the better" strategy to optimize the performance of $Sb_2Te$ photonic waveguide devices. Our subsequent experimental characterizations of $Sb_2Te$ thin films and optical device measurements fully verify our theoretical predictions. In particular, we reveal the unconventional optical properties of metastable crystalline $Sb_2Te$, and utilize these features for device design, yielding a simultaneous improvement in both the programming window and the optical loss. Overall, an optical programming precision exceeding 7-bit is achieved using a single waveguide cell, setting a new record for all-optical phase-change memory devices. Our work serves as a compelling example of computational material design, which demonstrates the predictive power of multiscale simulations in guiding the design of phase-change photonic devices for enhanced performance.

## 1. Introduction

The active development of artificial intelligence (AI) applications poses a major challenge to the classical von Neumann architecture, because massive volumes of data need to be shuttled back and forth between separated computing and storage units, giving rise to severe power consumption and latency issues. Resistive-switching devices with collocated computing and storage capabilities show great promise for energy-efficient AI computation[1-4]. Chalcogenide phase-change materials (PCMs) are one of the leading candidates for resistive-switching applications[5-8]. PCMs exploit the large contrast in electrical resistance or optical transmittance between their amorphous state (logic state = 0) and crystalline state (logical state = 1) to encode digital information. Both standalone and embedded phase-change electronic memories are commercially available in the market[9]. The contrast window in electrical resistance is also wide enough to accommodate multiple intermediate states for in-memory computing[10-12]. Furthermore, the rapid development of silicon photonics has unlocked new possibilities for PCM-based on-chip photonic computing[13-23]. The more transmittance levels a single photonic device can accommodate, the higher the efficiency and precision of photonic computing will be. However, this critical device parameter is still limited for PCM-based all-optical silicon-on-insulator (SOI) waveguides.

Density functional theory (DFT) and DFT-based ab initio molecular dynamics (AIMD) simulations, have played a central role in understanding the fundamental mechanisms of PCMs[24-27]. Combined theoretical and experimental investigations at the atomic scale have enabled the design of new PCMs with unprecedented properties[28-31], serving as typical examples of AI for materials. Based on thorough DFT/AIMD-generated materials databases, a couple of machine-learned interatomic potentials (MLIP) have been developed for various compositions on the Ge-Sb-Te ternary diagram[32-35]. The state-of-the-art MLIP-driven molecular dynamics simulations could support full-cycle device-scale atomistic simulations[36, 37]. Another pathway towards device-scale modeling is to use the DFT-calculated physical properties as input parameters for coarse-grained device simulations. For instance, we have performed finite-difference time-domain (FDTD) simulations for various PCM-based optical devices using DFT-calculated refractive indices, and proposed useful strategies for device optimizations[38-42]. It is highly desirable to experimentally verify some of these predictions in practical photonic devices. Guided by device modeling via multiscale simulations, experimental research on PCM-based photonic devices can be more targeted and efficient.

Growth-dominant PCMs, such as $Sb_2Te$ (ST) and Ag/In doped Sb-Te (AIST) alloys[43-45], which were initially exploited for rewriteable optical data storage[46], are gaining new attentions for on-chip photonic applications in miniaturized devices[47]. As the active volume reduces, crystallization operations of these materials can also be completed using short laser pulses, although their crystallization only proceeds via the amorphous–crystal boundaries. This interfacial growth process is not affected by the strong stochasticity inherent to statistical nucleation, and thus yields more predictable changes in optical properties upon step-wise crystallization[38]. Here, we performed multiscale simulations of ST and proposed a “the shorter the better” strategy to optimize the performance of ST-based photonic waveguide devices. Then we conducted thorough experimental investigations of ST thin films and all-optical waveguide devices under different annealing conditions to verify our theoretical predictions. We demonstrated the critical role of a metastable disordered crystalline state in widening the contrast window and reducing the optical loss of ST-based waveguide devices simultaneously. We achieved 158 distinct transmittance levels using standard waveguide devices, which sets a new record for all-optical phase-change photonic computing, while introducing no additional complexity to device engineering.

## 2. Results and Discussion

Figure 1a shows the multi-fold phase transition paths of ST, in which the models of the relevant phases were generated using DFT optimization and AIMD melt-quench calculations. The amorphous (a-) ST model contains 216 atoms in a cubic box with the edge length of 19.13 Å, corresponding to the experimental density of 6.34 g $cm^{-3}$ [48]. According to our previous AIMD calculations[38], a-ST transforms to a metastable (m-) rhombohedral structure with random occupation of the rocksalt lattice sites by Sb and Te atoms after rapid crystallization at ~600 K. Upon further chemical ordering (thermal annealing), the total energy of this m-state is gradually reduced until the ground (g-) state is reached, namely, a rhombohedral A7 structure with alternating $Sb_2Te_3$ QL and $Sb_2$ BLs. The g-ST model was built in a hexagonal unit cell of 9 atoms with $a$ = 4.27 and $c$ = 17.63 Å [49]. For m-ST, we constructed a 5×5×1 supercell of the ordered hexagonal unit cell, and randomly distributed Sb and Te atoms (in total 225 atoms) to the lattice sites with a Sb:Te ratio of ~2:1 per atomic layer. Then the atomic coordinates of the m-ST model were relaxed using DFT. Introducing the quantity $l_{Te}$, which represents the atomic concentration of Te atoms in the Te-rich layers, the transformation from m-ST ($l_{Te}$ ~ 33%) to g-ST ($l_{Te}$ = 100%) can be regarded as the ordering of Te atoms into specific layers until regularly stacked pure Sb/Te layers are formed. The full atomic structures of the four models are included in Figure S1 and the corresponding coordinates are openly available in the CAID repository (see Data Availability).

Next, we carried out electronic structure and optical response calculations using the Heyd–Scuseria–Ernzerhof (HSE06) hybrid functional[50], which yields results in better agreement with experimental data than the standard gradient-corrected functional[51]. Figure 1b shows the calculated density of states (DOS) for the three ST states. The a-ST model has a clear energy gap, and the g-ST model shows a pseudo gap, while the DOS of m-ST has a dip at the Fermi level. We calculated the dielectric functions of the three states (Figure S2) and converted them into the refractive index ($n$) and extinction coefficient ($k$) in the spectrum range of 400–2000 nm (Figure 1c), covering the visible-light range for non-volatile color display and the near-infrared range for photonic memory/computing. In comparison with the a-ST model, the g-ST model shows slightly smaller $n$ values below ~600 nm and systematically larger $n$ values at larger wavelengths. The $n$ curve of the m-ST model follows a similar trend, but crosses the one of a-ST at a much higher wavelength ~1250 nm. The overall contrast window between the a- and m-ST model is larger than that between the a- and g-ST model, particularly above ~1500 nm. Regarding $k$, the two crystalline models both show higher values than the amorphous model across the whole wavelength range. However, the contrast window between the a- and m-ST model is larger than that between a- and g-ST. These calculations suggest that the disordered crystalline state is more suitable for waveguide applications, as it offers a larger programming window.

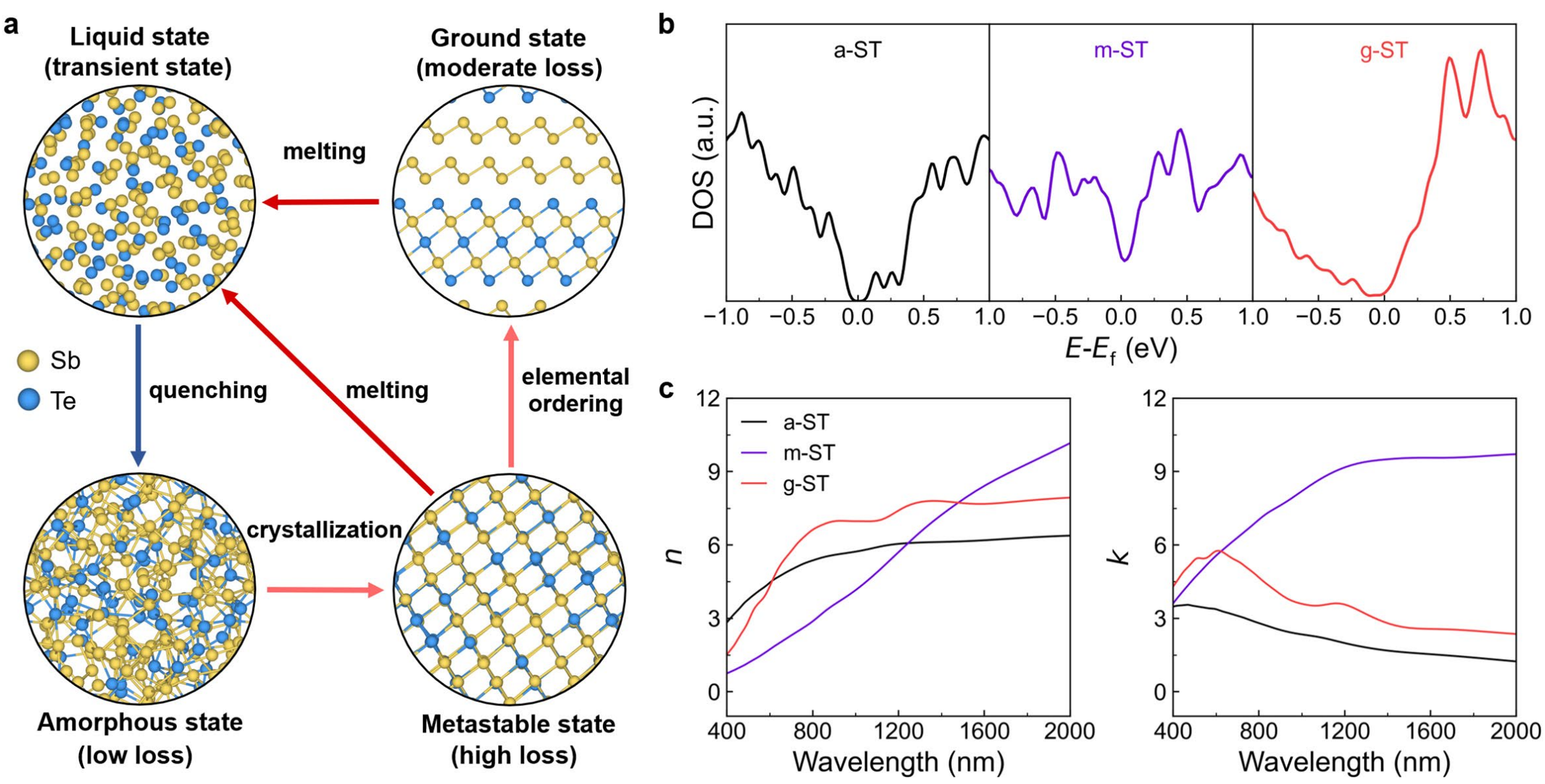

**Figure 1.** *Ab initio* calculations of liquid, amorphous and crystalline ST phases. **a** The multi-fold phase transitions between the liquid, amorphous, metastable crystalline and ground states of ST. **b** The calculated DOS and (c) refractive index *n* and extinction coefficient *k* of the three solid states using the HSE06 hybrid functional.

To verify this prediction experimentally, we prepared several ST films of ~100 nm thickness by magnetron sputtering. A ZnS:$SiO_2$ capping layer of ~20 nm was deposited to prevent potential oxidation. The thickness of our samples was determined by measuring the cross section under scanning electron microscope (SEM), and the compositional homogeneity was examined using energy-dispersive X-ray spectroscopy (EDS) mapping, as shown in Figure S3. The ST thin films were heated in an argon atmosphere to obtain crystalline samples. We first heated three ST samples to 160, 180 and 200 °C, which were then air-cooled down to room temperature. Next, we heated another three samples to 200, 250 and 300 °C with additional thermal annealing at these temperatures for 20 minutes before cooling. Figure 2a shows the X-ray diffraction (XRD) patterns of an as-deposited sample and the six thermally treated samples. The as-deposited film displays no visible XRD peaks except for the one from the Si substrate. For the annealed samples, there are three distinct peaks corresponding to the (004), (005) and (009) lattice planes of the ground state, indicating an out-of-plane texture of the films. With the increase of annealing temperature and/or annealing time, the peak positions gradually shift to lower diffraction angles.

The change in atomic structure is even more evident in the Raman spectroscopy measurements. Figure 2b shows a gradual blue shift and enhancement of the peak above 150 cm$^{-1}$, accounting for the vibrational $A_{1g}^{2}$ mode (155 cm$^{-1}$) of $Sb_2Te_3$ QL[52]. The peaks around 97 cm$^{-1}$ and 129 cm$^{-1}$ are related to the $E_g$ and $A_{1g}$ modes of Sb–Sb bonds[53], which display small variations for the six crystalline samples, because these bonds are always present regardless if the Te atoms are ordered.

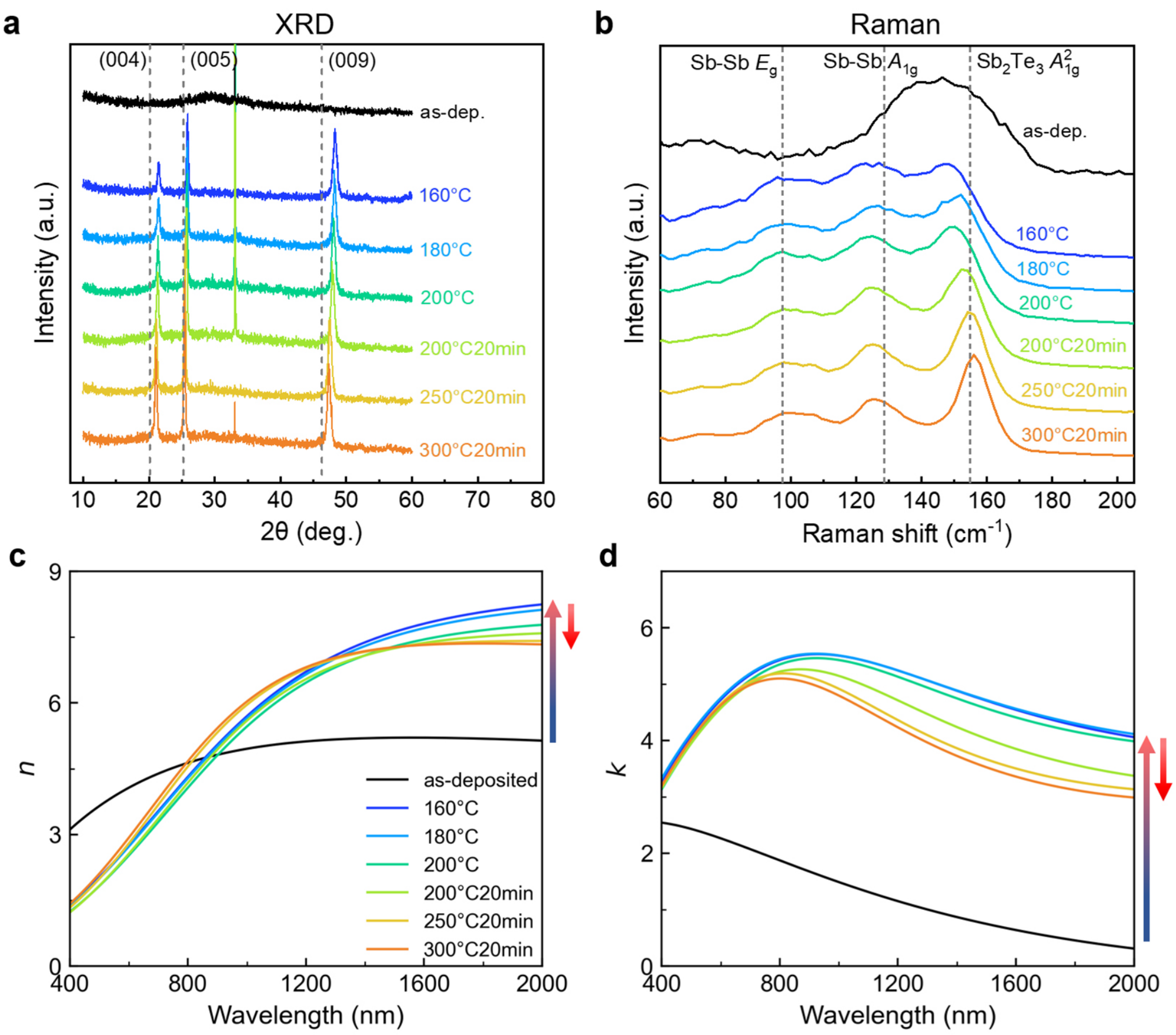

**Figure 2.** Structural and optical characterizations of ST thin films. **a** The XRD patterns measured for as-deposited and post-annealed ST thin films. The gray dashed lines indicate the positions of the XRD peaks identified from the standard card of $Sb_2Te$ (PDF#80-1722). **b** The Raman spectra measured for the ST thin films. The gray dashed lines indicate the Raman peaks for three typical vibrational modes. **c** The measured refractive index *n* and **d** the extinction coefficient *k* for ST thin films via spectroscopic ellipsometry experiments.

We next evaluated the impact of the structural transition on the refractive indices of ST thin films via spectroscopic ellipsometry measurements. As shown in Figure 2c and 2d, thermal annealing induces strong effects on the optical response, in particular, in the near infrared range. For instance, at 2000 nm, the *n* value increases from 5.1 to 8.2 upon crystallization, but gradually decreases upon further thermal annealing, and finally reaches 7.3 in the 300°C-20min sample. The same trend holds for *k*, which increases from 0.3 in the as-deposited sample to 4.1 in the 160°C sample, but then decreases to 3.0 in the 300°C-20min sample, as indicated by the colored arrows. In the whole near infrared range, the more disordered crystalline samples yield a bigger contrast window in *k* with respect to as-deposited amorphous sample than the more ordered crystalline samples. The difference in *k* for the crystalline samples becomes much smaller in the visible light range, consistent with our DFT prediction. As regards *n*, the c-ST samples show a major change from ~1.2 at 400 nm up to 8.2 at 2000 nm, while that of the a-ST sample changes from 3.1 at 400 nm to 5.1 at 2000 nm.

The crossing of the $n$ curves for the a-ST sample and the 160 °C c-ST sample takes place at 900 nm, and the crossing point gradually shifts to 800 nm upon thermal annealing of the crystalline samples. These two crossing points fall within the broader range determined computationally by considering the fully disordered and fully ordered crystalline ST states (Figure 1c). Despite potential numerical differences between theoretical values and optical measurements, this comparison indicates that the six crystalline samples are neither fully disordered nor fully ordered. This prediction is fully confirmed by the element-resolved atomic imaging experiments discussed below.

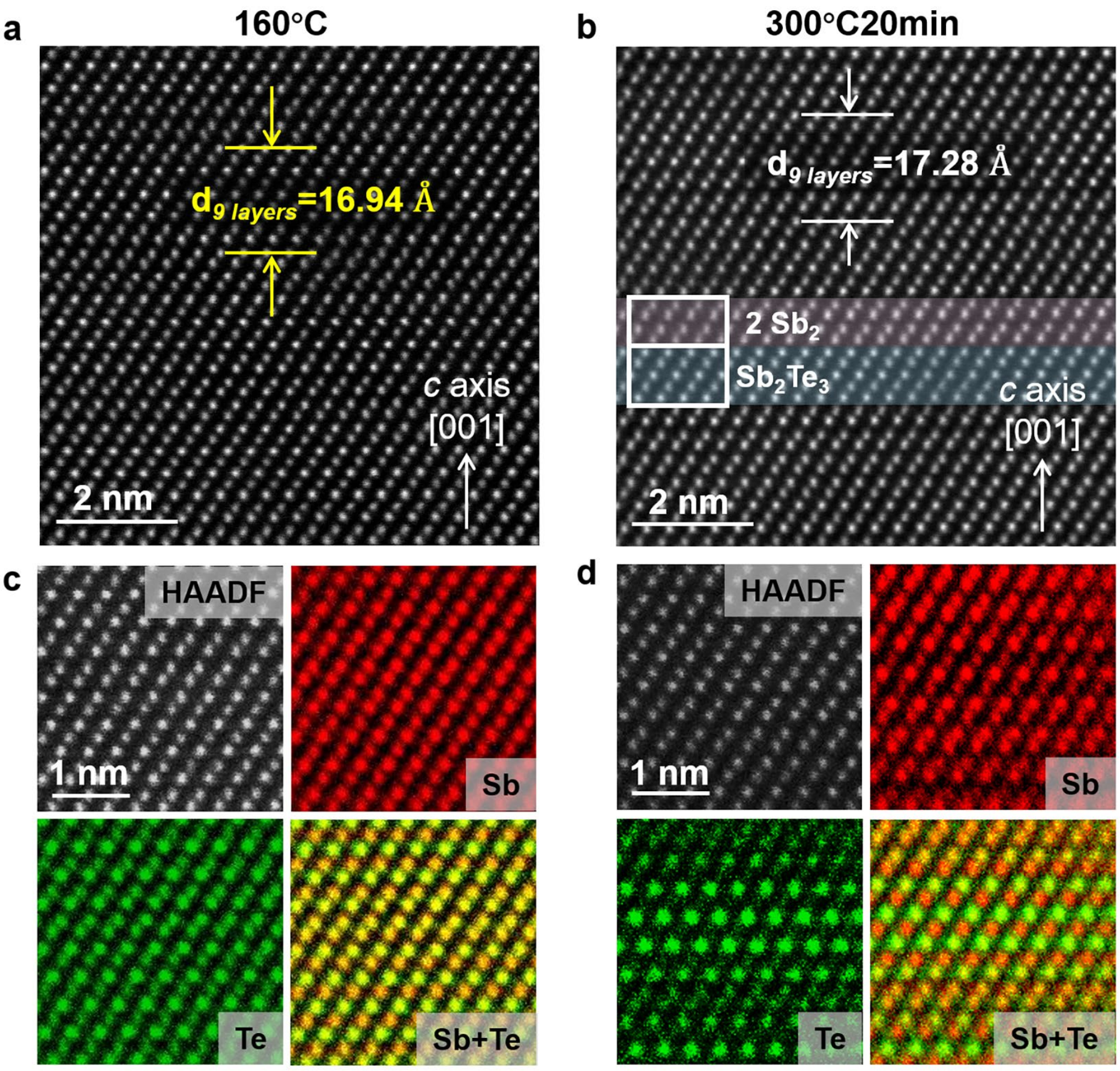

**Figure 3.** Atomic-scale structural characterizations of c-ST thin films. The atomic-scale HAADF-STEM images of **a** the 160°C c-ST sample and **b** the 300°C-20min c-ST sample. The corresponding atomic EDS mappings of **c** the 160°C c-ST sample and **d** the 300°C-20min c-ST sample.

We performed spherical aberration corrected high-angle annular dark-field scanning transmission electron microscopy (HAADF-STEM) and atomic EDS mapping experiments on the most disordered (160°C) and the most ordered (300°C-20min) c-ST samples. The layer spacings in the 160°C sample seem uniform (Figure 3a), while the structural features of the ordered A7 structure are more clearly visible in the 300°C-20min sample, where alternately stacked Sb-rich BLs and Te-rich QLs can be distinguished, as indicated by the pink and blue shaded areas (Figure 3b). The measured spacing for 9 layers along the [001] direction is also slightly enlarged in the ordered state. Atomic EDS mappings revealed small differences in Sb/Te concentration between adjacent layers in the 160°C sample (Figure 3c). In 300°C-20min sample, the ordering of Te atoms in specific layers is evident (Figure 3d). Nevertheless, this state is still far from the ideal structure, where Te atoms solely occupy the three atomic layers of one $Sb_2Te_3$ QL block. This elemental ordering process in Sb-Te crystals was also demonstrated via the state-of-the-art in situ STEM experiments by Zheng *et al.*[54].
Since the formation of a more ordered phase results in a reduction in the optical window, it is important to avoid additional heating for ST-based optical devices. This trend is opposite to the extensively studied $Ge_2Sb_2Te_5$ (GST)-based photonic devices[13-18], in which the vacancy ordering-induced phase transition from the rocksalt metastable state to the trigonal ground state[55-57] widens the optical window[58]. Therefore, a different strategy needs to be adopted for the fabrication and programming of ST-based devices. To provide detailed guidance, we carried out FDTD simulations of SOI waveguide memory devices based on the measured $n$ and $k$ profiles of the ST thin films. Figure 4a shows a typical waveguide device setup, where the thickness and width of the waveguide, and the thickness and length of the ST thin films and the indium tin oxide (ITO) capping layer, are denoted as $h_{wg}$ and $w_{wg}$, $h_{ST}$, $d_{ST}$, $h_{ITO}$ and $d_{ITO}$, respectively. We considered a 220-nm-thick SOI waveguide in a shallowly etched geometry covered with two thin-film layers on top, i.e., $h_{wg}$ = 0.15 μm, $w_{wg}$ = 0.45 μm, $h_{ST}$ = 10 nm and $h_{ITO}$ = 15 nm. We set the ST and ITO lengths to $d_{ST}$ = $d_{ITO}$ = 4, 2 or 1 μm for optical programming. The transmittance of light ($T$) for the ST-based waveguide devices can be calculated based on the input and output power ($P_1$ and $P_2$, respectively) as $T = P_2/P_1$. The wavelength of the incident light ranges from 1500 to 1600 nm. The measured $n$ and $k$ data in the range between 1535 nm and 1565 nm of the measured amorphous and crystalline $Sb_2Te$ thin films (black, blue and orange curves in Fig. 2c and 2d) were used as the input parameters for the FDTD simulations of the waveguide transmittance spectra in Fig. 4b. These $n$ and $k$ data were measured at room temperature for the as-deposited amorphous thin film and for the thin films crystallized at 160°C and 300°C, respectively.

Figure 4b shows the simulated spectra of transmittance for the three waveguide devices with different $d_{ST}$ operated in the three different ST states. By decreasing $d_{ST}$ from 4 μm to 1 μm, the device transmittance in the as-deposited a-ST state is increased from 42.8% to 80.3% due to weaker light absorption by the ST thin film. For the device with $d_{ST}$ = 4 μm, the overall transmittance window $\Delta T$ is 42.6%, and there is nearly no difference in transmittance when the device is programmed to the more ordered or the more disordered crystalline state. As expected, the impact of Te ordering becomes visible in shorter devices with $d_{ST}$ = 2 μm or 1 μm, and the $\Delta T$ between the more ordered and the more disordered c-ST state is increased to ~4.3% in the latter device (blue and orange lines in Figure 4b). The overall amorphous-to-crystalline transmittance window $\Delta T$ is enlarged to ~62.5% in the two devices. Figure 4c and 4d show the simulated electric field intensity $|\boldsymbol{E}|$ of the ON and OFF state, representing the largest transmittance contrast, at 1550 nm for the three devices. For the 1-μm device, it is possible to obtain an even larger contrast window if the device can be programmed to a more disordered crystalline state using ns laser pulses, as indicated by the red arrow in Figure 4b. It is also feasible to use DFT-calculated $n$ and $k$ data as input parameters for the FDTD simulations, which yields consistent results (Figure S4).

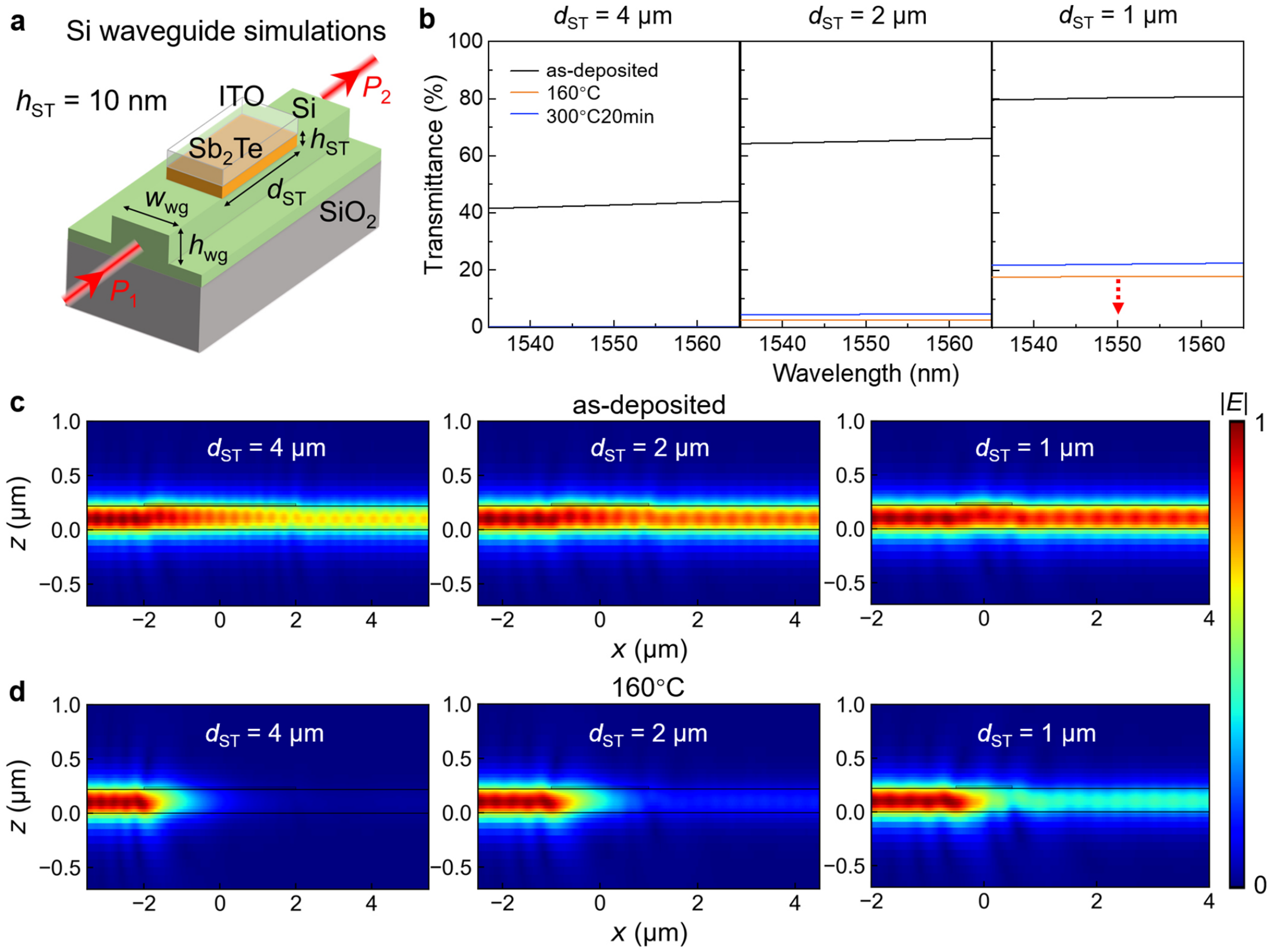

**Figure 4.** FDTD simulations of ST-based waveguide devices. **a** A sketch of PCM-based waveguide device. **b** The simulated transmittance spectra of ST-based waveguide devices with fixed film thickness $h_{ST}$ = 10 nm and varying length $d_{ST}$ = 4, 2 or 1 μm. The simulated electric field intensity |***E***| for the ST-based devices at a wavelength of 1550 nm using the measured *n* and *k* of **c** the as-deposited thin films and **d** the thin films heated to 160°C.

Following the atomistic understanding of ST, we fabricated SOI waveguides with short ST films of $d_{ST}$ = 2 and 1 μm. The passive SOI waveguides were manufactured by the silicon photonic foundry (CUMEC) using the multi-project wafer (MPW) service with 180 nm Complementary Metal Oxide Semiconductor (CMOS) node technology. In a post-fabrication process, a partial area (yellow region in Figure 5a) of the silicon dioxide upper cladding of the SOI waveguide was etched away for facilitating the ST deposition. Electron beam lithography was performed to open windows in a positive-tone resist (PMMA) for ST deposition. A patch-shaped ST thin film was integrated with the SOI waveguide by magnetron sputtering and resist removal. Then the transmittance of the waveguide devices was measured via the fiber-to-chip characterization setup sketched in Figure 5b. Optical pump pulses were generated by modulating a pump laser using an electro-optic modulator. The pulse power was then amplified by an erbium-doped fiber amplifier. A low-power continuous-wave laser served as a probe to measure transmittance during the all-optical switching process. Optical signals were coupled into and out of the silicon waveguides via a fiber array aligned with the on-chip grating couplers. Pump and probe light propagated in opposite directions along the waveguide. The probe light was dropped by a fiber circulator and an optical filter, and its signal was monitored in real time by a photodetector. More details about the measurement setup can be found in the Methods Section.

In these all-optical pump-probe experiments, ST thin films can be switched back and forth between the amorphized and recrystallized states by nanosecond amorphization (write) and crystallization (erase) optical pulses. Prior to optical switching, waveguide devices are usually annealed to form a fully crystalline state to allow better programming consistency. For the 1-μm device, the transmittance value changes from ~76% at 1530 nm to ~86% at 1570 nm in the as-deposited state, drops to ~21% in the 160°C state, and increases to ~35% in the 300°C-20min state across the measured wavelength range (Figure 5c). Clearly, the Te ordering leads to an increase in transmittance by ~14% in this device. Regarding the 2-μm device, it shows a similar trend but smaller transmittance increase (~6%) induced by Te-ordering (Figure 5c). These observations are consistent with our theoretical predictions.

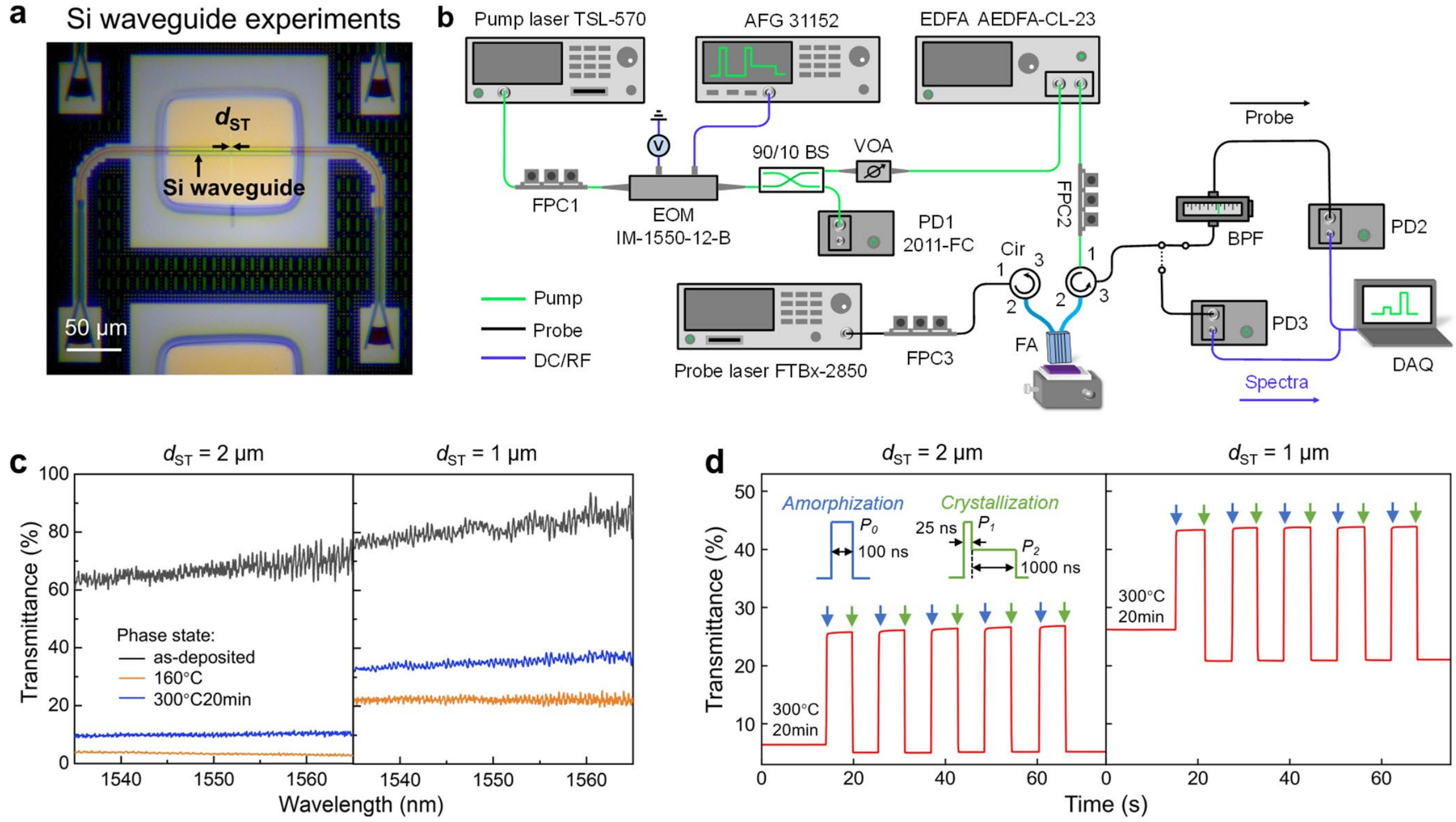

**Figure 5.** Optical experiments of ST-based waveguide devices. **a** The optical microscope image of the waveguide device. **b** The all-optical pump-probe experimental setup. AFG: arbitrary function generator, EOM: electro-optic modulator, EDFA: erbium-doped fiber amplifier, FPC: fiber polarization controller, BS: beam splitter, VOA: variable optical attenuator, Cir: circulator, PD: photodetector, FA: fiber array, BPF: bandpass filter, DAQ: data acquisition, DC: direct current, RF: radio frequency. **c** The measured transmittance spectra of waveguide devices with $d_{ST}$ = 2 or 1 μm in the as-deposited state and two post-annealed states. **d** The transmittance response of the two waveguide devices upon optical switching at nanosecond timescale. The shapes and widths of the programming pulses are sketched as insets, where $P_0$, $P_1$ and $P_2$ are 12.17, 11.87 and 3.84 mW for $d_{ST}$ = 2 μm, and 10.72, 10.53 and 3.19 mW for $d_{ST}$ = 1 μm, respectively.

Starting with a well-annealed crystalline state, a train of amorphization and crystallization optical pulses was sent to program the 2-μm device and the 1-μm device (Figure 5d). We adopted the programming pulse scheme with a single-step pulse for melt-quenched amorphization and a double-step pulse for crystallization[14] (Figure 5d inset). The pulse power and width were properly adjusted to achieve reliable and reversible optical switching.

Blue and green arrows indicate amorphization and crystallization events, respectively. The ST devices were switched back and forth for five consecutive cycles. After the first amorphization pulse, the transmittance was increased to 25.7% and 43.3% in the 2-μm and 1-μm device, much smaller than that of the as-deposited state. This is because only part of the device was switched due to local heating via photothermal effects, and the rest of the device was still in the well-annealed crystalline state with relatively high optical loss. Subsequent crystallization of this melt-quenched amorphous volume resulted in a larger contrast window, indicating that a more disordered crystalline state was obtained as compared to the initial crystalline state. This trend is more evident in the 1-μm device. After an amorphization pulse, we heated the device to 160 °C and carried out optical measurement after natural cooling. This state showed a higher transmittance value than the recrystallized state obtained via optical pulsing (Figure S5), showing that very fast crystallization can indeed result in an even more disordered crystalline state. Overall, the 1-μm device shows a slightly larger optical window of 22.5% (20.6% for the 2-μm device) and smaller optical loss in both ON and OFF states, and is therefore more suitable for multi-bit programming in a scalable crossbar array.

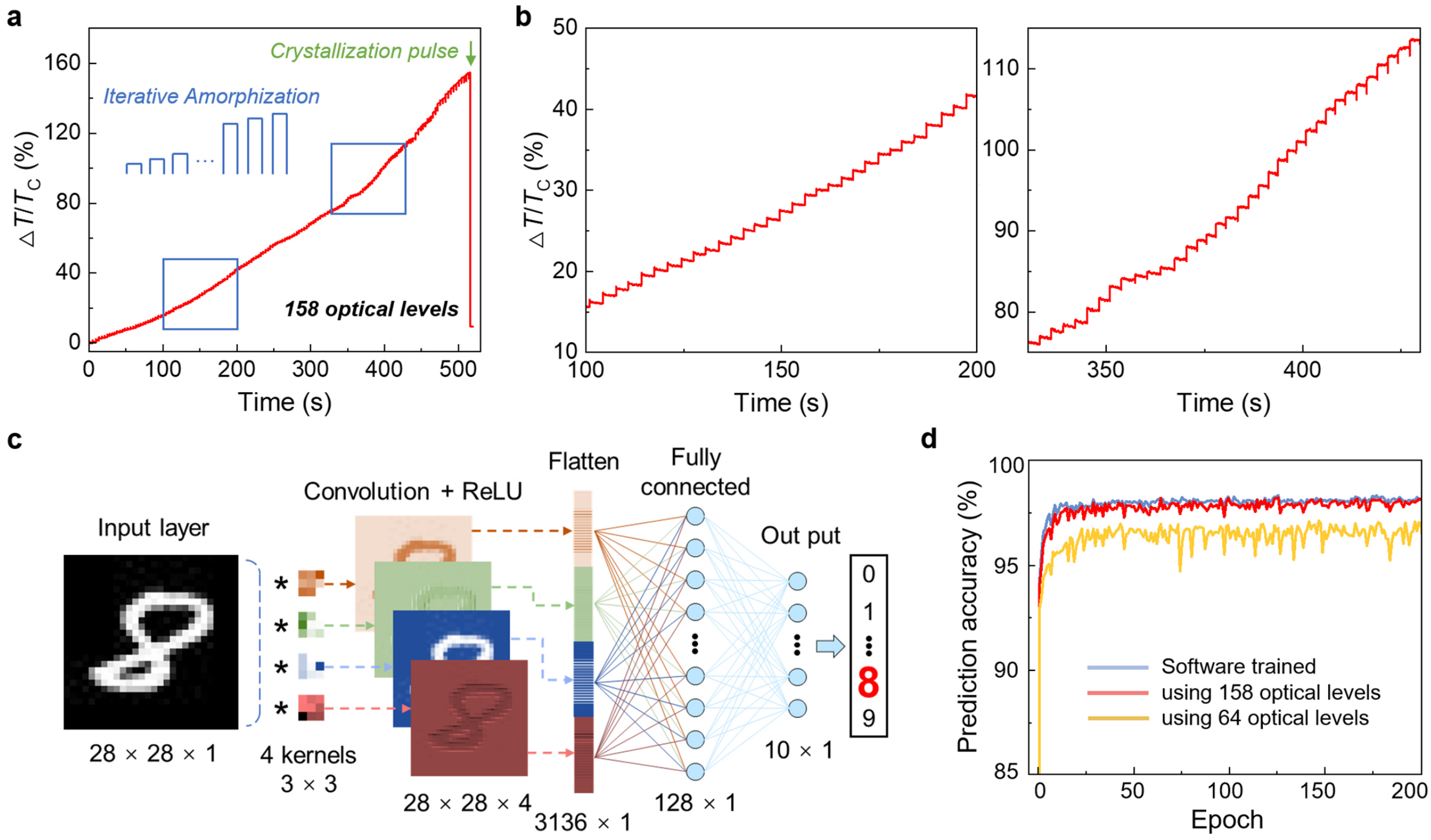

**Figure 6.** Performance of ST-based waveguide devices and predicted recognition accuracies. **a** Experimental demonstration of multilevel programming capacity (158 levels) of the ST-based waveguide device ($d_{ST}$ = 1 μm). **b** Two zoom-in images of the areas marked by the blue boxes in a. **c** The layer structure of an artificial neural network for digit recognition. (d) The predicted accuracy of the neural network using 158 and 64 optical levels.

To assess the multilevel capability of ST, we performed iterative amorphization operations on the 1-μm device. A variable optical attenuator was adopted to control the pump power of amorphization pulses. As shown in Figure 6a inset, a sequence of pulses with monotonously increased pulse power was sent to progressively enlarge the amorphization area of the 1-μm device. The switching contrast is defined by $\Delta T/T_c$, where $\Delta T=T_a-T_c$, $T_a$ and $T_c$ are transmittance values of the ST device for the amorphized and crystallized states, respectively. Figure 6a shows 158 optical levels; these discrete states are better discerned in the zoom-in images shown in Figure 6b. It is noted that at the last stage of iterative amorphization, larger pump power was needed, which resulted in a transient decrease in transmittance after sending a partial amorphization pulse. This behavior is due to the temporary increase in local temperature that leads to a slightly larger optical loss of PCM. After thermal dissipation, the device was programmed to a higher transmittance state in a non-volatile manner.
With fine-tuning of the crystalline–amorphous volume ratio, our ST device enables the largest number of programmable optical levels on SOI waveguide devices via all-optical approach, far more than the 64 and 45 levels obtained using GST [18] and AIST [47] all-optical waveguide devices. The device-to-device variability was tested, and the high amount of transmittance levels was well reproduced. We also performed the multilevel programming operation over five times using one ST-based device, and the multiple transmittance levels can still be well distinguished with inclusion of error bars. Besides, we prolonged the recording time of the high- and low-transmissive levels and observed limited variations. The wide contrast window can also be sustained upon repetitive write/erase cycling. These device testing data can be found in Figures S6-S9.

More optical levels per device could lead to a higher programming precision for photonic synapse emulation, and our ST device can support a >7-bit programming precision per node for photonic neural networks. As an attempt, we carried out numerical simulations of a convolutional neural network (CNN) using the measured transmittance values of the 1-μm device. Figure 6c shows the CNN framework, which involves a convolutional layer containing four 3×3 convolutional kernels, a ReLU nonlinear activation layer, a flatten layer, and three fully-connected layers. We used the MNIST handwritten digits database for training the synaptic weight banks based on the stochastic gradient descent and standard backpropagation algorithms[59]. The detailed training process with prediction accuracy versus training epoch is shown in Figure 6d. Our ST devices-based CNN with 158 optical levels per node could potentially support an inference accuracy of ~98% (red curve), reaching the level of software-trained CNN (blue curve). We also used the optical data of GST devices with 64 levels per node for the same CNN simulation, but the prediction was less accurate and was subjected to higher randomness (yellow curve).

We note that this fabrication strategy is different from the standard device miniaturization approach, because the waveguide devices were produced using the common 180 nm CMOS technology node in foundry. Hence, the size of waveguide devices was not reduced. The only difference was that we proposed to change the material length of PCM thin film from several μm to 1 μm using electron beam lithography, adding marginal complexity to the fabrication process. We further estimate the scalability of $Sb_2Te$ in a potential crossbar array by considering the insertion loss of waveguide cells and the light splitting ratio of the array (see Figure S10). Our estimated array size is ~45×45 with a total loss of −40 dB, which is comparable to those estimated in a GST metasurface based mode converter array [60]. Yet, our approach that utilizes the metastable phase of $Sb_2Te$ offers a more straightforward way for large-scale integration. We also note that by integrating additional spatial optical pump sources[61] or embedding electrically pulsed microheaters[62], larger areas of a single PCM waveguide cell can be programmed to achieve high programming precisions[61-65]. Although these approaches lead to higher manufacturing costs, they can enable simultaneous programming and optical computing. Hence, it is also anticipated to test the performance of $Sb_2Te$ in these setups.

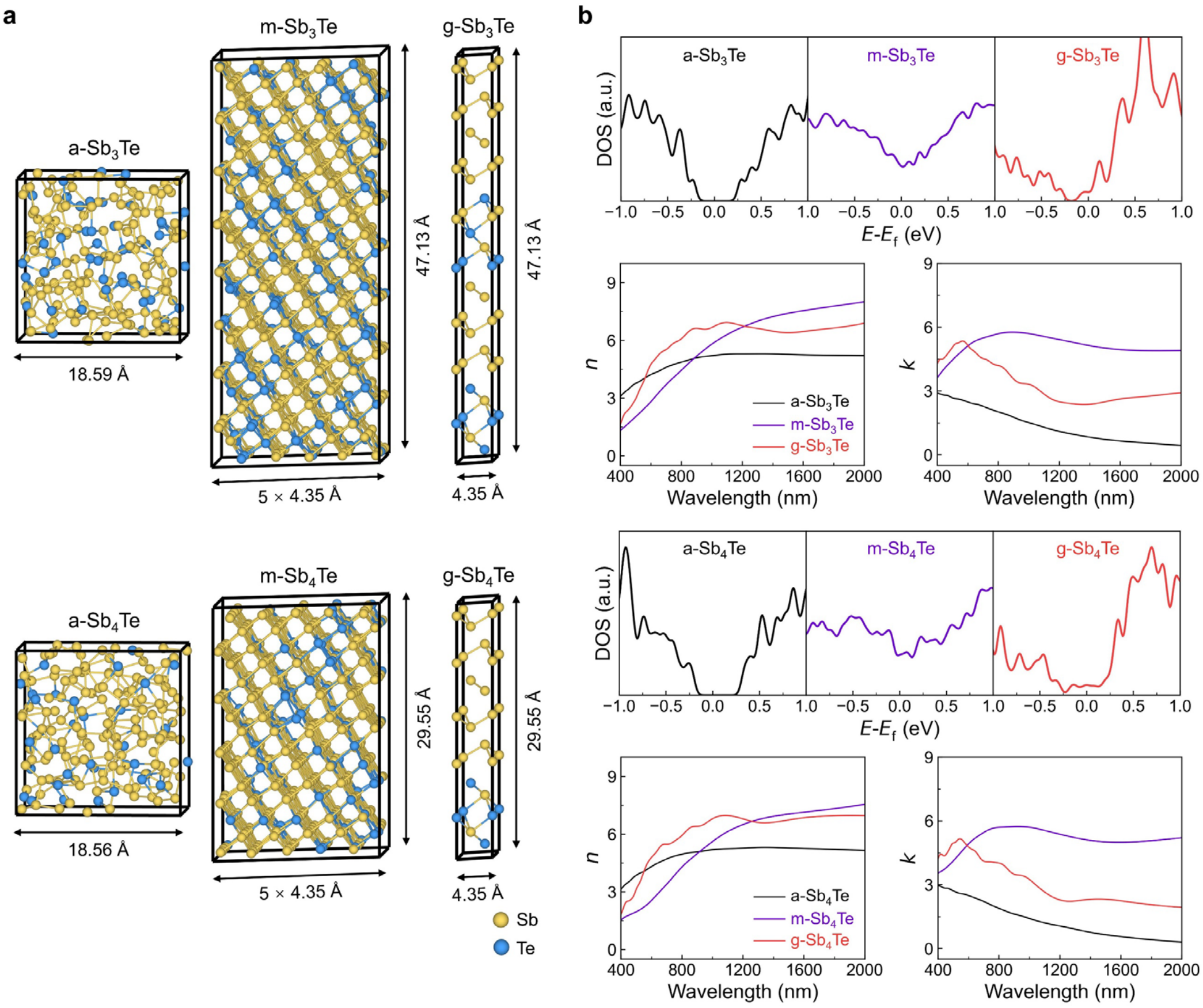

**Figure 7.** Electronic structure and optical calculations of $Sb_3Te$ and $Sb_4Te$. **a** The atomic structure of a-$Sb_3Te$, a-$Sb_4Te$, g-$Sb_3Te$, g-$Sb_4Te$, m-$Sb_3Te$ and m-$Sb_4Te$. **b** The DFT-calculated DOS, *n* and *k* calculated with HSE06 hybrid functional.

As discussed above, our *ab initio* predictions of the optical properties are well confirmed by our optical measurements. This atomic understanding helps us optimize the performance of $Sb_2Te$-based all-optical waveguide devices. We note that the presence of metastable crystalline phase and the subsequent elemental ordering effects should be a generic feature for Sb-Te alloys. The common compositions also include $Sb_3Te$ and $Sb_4Te$. Their ground state forms a similar A7 rhombohedral structure with one $Sb_2Te_3$ QL block but more $Sb_2$ BLs along the *c* axis in the unit cell. Figure 7 shows the atomic models, DOS, and optical profiles of amorphous, metastable and ground state models of $Sb_3Te$ and $Sb_4Te$, which were calculated using HSE06 hybrid functional. The amorphous and ground state models were taken from our previous work[66]. The 5×5×1 supercell models of m-$Sb_3Te$ and m-$Sb_4Te$ were built based on their ground state models, and the lattice sites were randomly occupied by Sb and Te atoms with a Sb:Te ration of ~3:1 and ~4:1 per atomic layer. The atomic coordinates were relaxed with fixed lattice edges. The DOS profiles of $Sb_3Te$ and $Sb_4Te$ show a similar trend as that of $Sb_2Te$, i.e., the energy bands around the Fermi level are much more filled in the disordered metastable crystalline state than in the ordered ground state. This electronic structure feature would give rise to a larger optical window between amorphous and metastable crystalline phases than that between amorphous and rhombohedral phases.

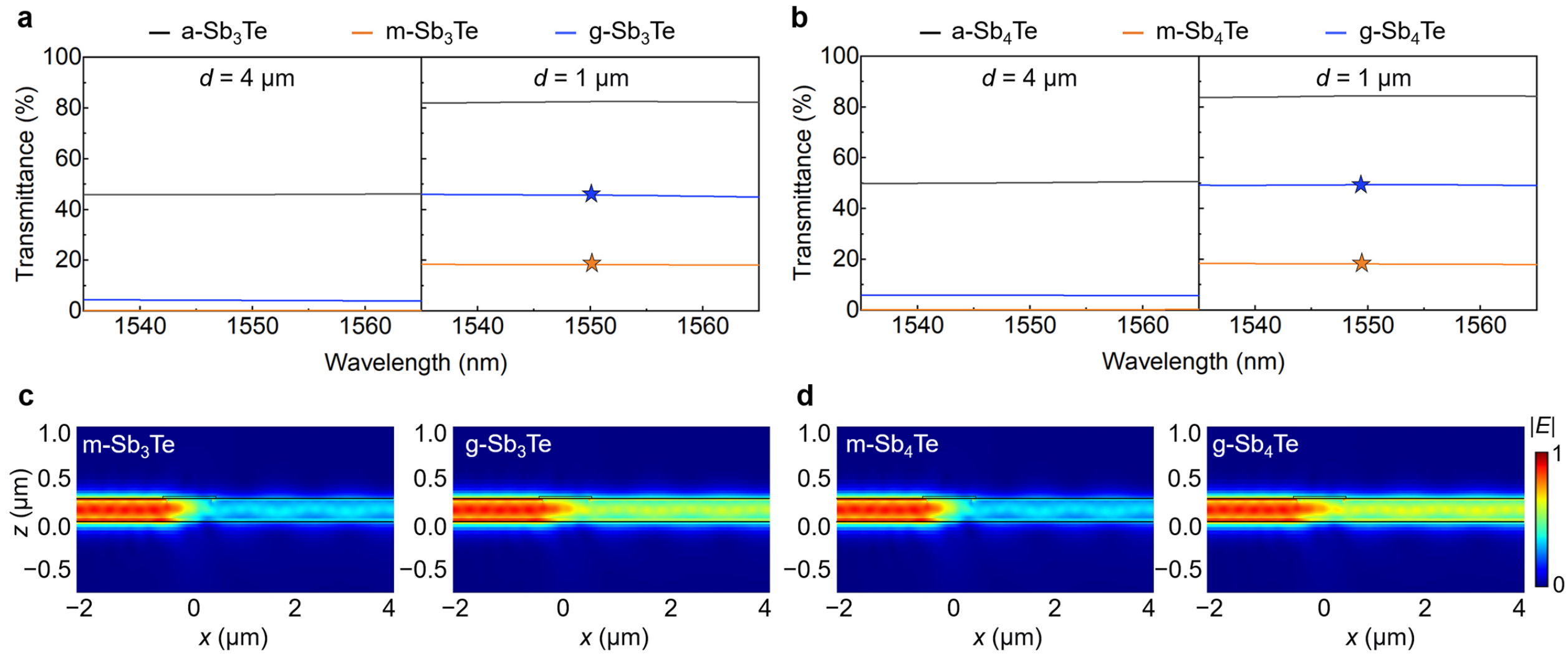

**Figure 8.** FDTD simulations of $Sb_3Te$- and $Sb_4Te$-based waveguide devices. **a** The simulated transmittance spectra of **a** $Sb_3Te$-based and **b** $Sb_4Te$-based waveguide devices with fixed film thickness of 10 nm and varying length $d$ = 4 or 1 μm. The simulated electric field intensity $|\boldsymbol{E}|$ for the **c** $Sb_3Te$ and **d** $Sb_4Te$-based devices in the m- and g-phases at a wavelength of 1550 nm (as marked by blue and orange stars in a and b) using the DFT-calculated $n$ and $k$ values.

As shown in Figure 8, our FDTD simulations support this claim. We used the same device setup as illustrated in Figure 4. Due to the lack of experimental $n$ and $k$ data as a function of annealing temperature, we took the DFT-calculated $n$ and $k$ values of the six $Sb_3Te$ and $Sb_4Te$ models as input parameters for these FDTD simulations. For the two 4-μm devices, the simulated transmittance values are very low in both the g- and m-phases, and the difference between the two spectra is small. As the material length $d$ is reduced to 1 μm, all three phases of the $Sb_3Te$- and $Sb_4Te$-based devices are more transmissive, and the programming window between the m- and a-phases is much larger than that between the g- and a-phases (Figure 8a and 8b). In fact, the transmittance contrast between the m-phase and the g-phase is already sufficiently large to represent the OFF and ON states, as highlighted by the simulated electric field intensity $|\boldsymbol{E}|$ shown in Figure 8c and 8d. This is because these two phases correspond to the fully disordered and fully ordered crystalline states of $Sb_3Te$ and $Sb_4Te$, respectively. Experimentally, it is more likely to obtain intermediate states with medium-level of Te ordering (e.g., Figure 3) and relatively smaller contrast. Nevertheless, we note that all the three Sb-Te compositions follow the same optical trend. To enhance multilevel programming capacity while reducing optical loss, the optimization of these growth-driven PCMs should follow the same "the shorter the better" design strategy. Besides, additional temperature rise should be avoided to prevent elemental ordering effects.

## 3. Conclusion

In summary, we provided experimental evidence via optical measurements for an unconventional change in optical properties in $Sb_2Te$-based thin films and waveguide devices. The refractive index and extinction coefficient of $Sb_2Te$ thin films in the near-infrared range first increased upon crystallization into the metastable disordered rhombohedral phase and then decreased upon further thermal annealing. The ordering of Sb and Te atoms into layered structures resulted in an opposite change in optical response as compared to the vacancy ordering process in crystalline GST alloys. Our combined theoretical and experimental study at the atomic scale helped us design and optimize waveguide devices with a simultaneous improvement on the programming window and the optical loss. In total, 158 distinguishable transmittance levels and 156% switching contrast, which are one of the highest values compared with those reported for all-optical waveguide devices (see Table S1), were achieved using a 1-μm SOI waveguide device via the all-optical approach. It supports a CNN simulation with high accuracy for the classification of the images of the MNIST database. The improved device performance by tailoring properties of PCM adds no complexity to the manufacturing and integration of waveguide devices in crossbar arrays. Our further multiscale simulations predicted the elemental ordering effects to be a generic feature for the Sb-rich Sb-Te alloys. Taken together, our work showcases how the scheme of AI for materials can guide the discovery of unconventional metastable states that have major implications for the design and operation of phase-change photonic devices.

## 4. Methods

*Ab initio calculations*: The melt-quenched amorphous $Sb_2Te$ model was generated via AIMD calculations using the CP2K package[67], and the density of this model was adjusted to the experimental value of amorphous $Sb_2Te$ thin film[48]. Structural relaxation and dielectric function calculations of the amorphous models and the two crystalline models were made using the VASP Package[68]. For relaxation, the Perdew–Burke–Ernzerhof (PBE) functional[51] and projector augmented-wave (PAW) pseudopotentials[69] were used with an energy cutoff of 400 eV. The $\Gamma$ point was used to sample the Brillouin zone of amorphous and disordered crystal models, and a $k$-point mesh of 12×12×2 was used for the ordered crystal model. The electronic structure and dielectric calculations were performed using the HSE06 hybrid functional[50], and the $k$-point mesh was increased to 24×24×5 for the ordered crystal model for better convergence of optical data. The dielectric calculations done within the independent particle approximation neglecting local field effects and many body effects, which was proven to be adequate to characterize the optical properties of PCMs[70-72].

*FDTD simulations*: The photonic waveguide devices were modelled by 3D FDTD (Lumerical FDTD Solutions, https://www.ansys.com/products/optics/fdtd). The incident light is with the fundamental transverse magnetic (TM) mode and a wavelength range of 1500–1600 nm. The light source and the monitor were located at different sides of the PCM film to detect transmittance spectra. The perfectly matched layer (PML) boundary condition was applied to all boundaries. The grid sizes were set fine enough to obtain converged simulation results, with 1 nm grid size for the $Sb_2Te$ and ITO films along the thickness direction. The optical constants are assumed to be spatially uniform across the phase-change programming region.

*Sample preparation*: $Sb_2Te$ amorphous films with thickness of ~100 nm were deposited by magnetron sputtering on a Si(100) substrate using a stoichiometric $Sb_2Te$ alloy target (20 W power from a direct current voltage source, 30 sccm argon flow rate). The films were then treated with six different thermal conditions. For three of the thermal conditions, samples were heated up to 160°C, 180°C and 200°C, respectively, and then naturally cooled to ambient temperature. For the other three, samples were annealed at 200°C, 250°C and 300°C respectively for 20 minutes. For all samples, the heating rate was 10 $K{\cdot}min^{-1}$, and the thermal process was done in argon atmosphere.

*Structural and compositional characterizations*: The thickness of the as-deposited films was determined by measuring the cross section using a Hitachi SU8230 scanning electron microscope (SEM). Using energy-dispersive X-ray spectroscopy (EDS), the concentration of Sb and Te was determined to be 69.04 and 30.96 at%, respectively. The X-ray diffraction (XRD) measurements were performed using a Bruker D8 ADVANCE diffractometer with Cu $K_\alpha$ radiation source. The Raman spectra were collected using a Renishaw invia Qontor Raman microscope with a 532-nm laser at ambient temperature. Specimens for transmission electron microscopy (TEM) characterization were prepared by Hitachi NX5000 focused ion beam system with a Ga ion beam at 30 keV beam energy and polished at 5 keV. High-angle annular dark field scanning transmission electron microscopy (HAADF-STEM) imaging and atomic EDS mappings were performed on a JEM-ARM300F2 microscope with double spherical aberration correctors.

*Optical experiments and fitting*: Spectroscopic ellipsometry measurements were performed with a UVISEL PLUS ellipsometer. The incidence angle was set to 70° with the light source of xenon lamps. The spectra data were fitted using the CODE software (www.mtheiss.com). The refractive indices of $Sb_2Te$ were obtained by fitting the measured spectra with a multi-layer model involving the substrate, the $Sb_2Te$ film and the capping layer. The dielectric model of amorphous $Sb_2Te$ included a constant dielectric background and a Tauc–Lorentz oscillator[73] describing inter-band transitions, while an additional Drude contributor for free carrier absorption was incorporated for the crystalline films. The dielectric functions of the substrate and the capping layer were determined independently based on reference samples. Such method has been established in optical measurements of typical phase-change materials[74].

*Device measurement setup*: Two tunable lasers (TSL-570, Santec and FTBx-2850, EXFO) are used as the pump and probe lasers respectively, with wavelengths of 1550 nm and 1540 nm, and power of 1 mW and 10 mW. To generate ns-level write and erase pulses, the pump laser beam first passed through a fiber polarization controller, and was next modulated by an electro-optic modulator (IM-1550-12-B, Optilab) driven by an arbitrary function generator (AFG31152, Tektronix) to send out write and erase pulses. An electronic variable optical attenuator (VOA) (V1550A, THORLABS) was programmed to control the write pulse power. Pump pulse power was further amplified by an erbium-doped fiber amplifier (EDFA) (AEDFA-CL-PS-23-B-FA, Amonics Ltd.). Pump and probe were coupled into the SOI photonic integrated circuits from different input ports based on the grating couplers with optimized transmission using FPCs, such that pump and probe transmitted along the opposite directions through the photonic memory cells. Noises from pump reflection were filtered by an optical band-pass filter (optical tunable filter-320, Santec) for the probe detection. A 200 kHz low-noise photoreceiver (2011-FC, Newport) was used to record dynamic change of the probe transmission.

## Supporting Information

Supporting Information associated with this article will be available online.

## Acknowledgments

The authors thank the support of National Natural Science Foundation of China is acknowledged (grant number 62374131 for W.Zhang, 62405242 for W.Zhou). W.Z. thanks the support of the 111 Project (B25007). The authors acknowledge the support of the HPC platform of XJTU and the Computing Center in Xi'an for computational resources. The authors acknowledge Chaobin Zeng from Hitachi High-Tech Scientific Solutions (Beijing) Co., Ltd. and Yuanbin Qin for technical support on STEM experiments. The authors also thank Jia Liu at Instrument Analysis Center of XJTU for the assistance regarding Raman spectroscopy experiments. The International Joint Laboratory for Micro/Nano Manufacturing and Measurement Technologies of XJTU is acknowledged. R.M. acknowledges funding from the PRIN 2020 project “Neuromorphic devices based on chalcogenide heterostructures” funded by the Italian Ministry for University and Research (MUR).

## Conflict of Interest

The authors declare no competing interests.

## Data Availability Statement

Upon formal journal publication, the data that support the findings of this study will be openly available in the CAID Repository at https://caid.xjtu.edu.cn/info/1003/1953.htm, reference number 1953.

# Supporting Information

Multiscale simulations guided advances for all-optical phase-change waveguides

*Hanyi Zhang, Wanting Ma, Wen Zhou, Xueqi Xing, Junying Zhang, Tiankuo Huang, Ding Xu, Xiaozhe Wang, Riccardo Mazzarello, En Ma, Jiang-Jing Wang, Wei Zhang*

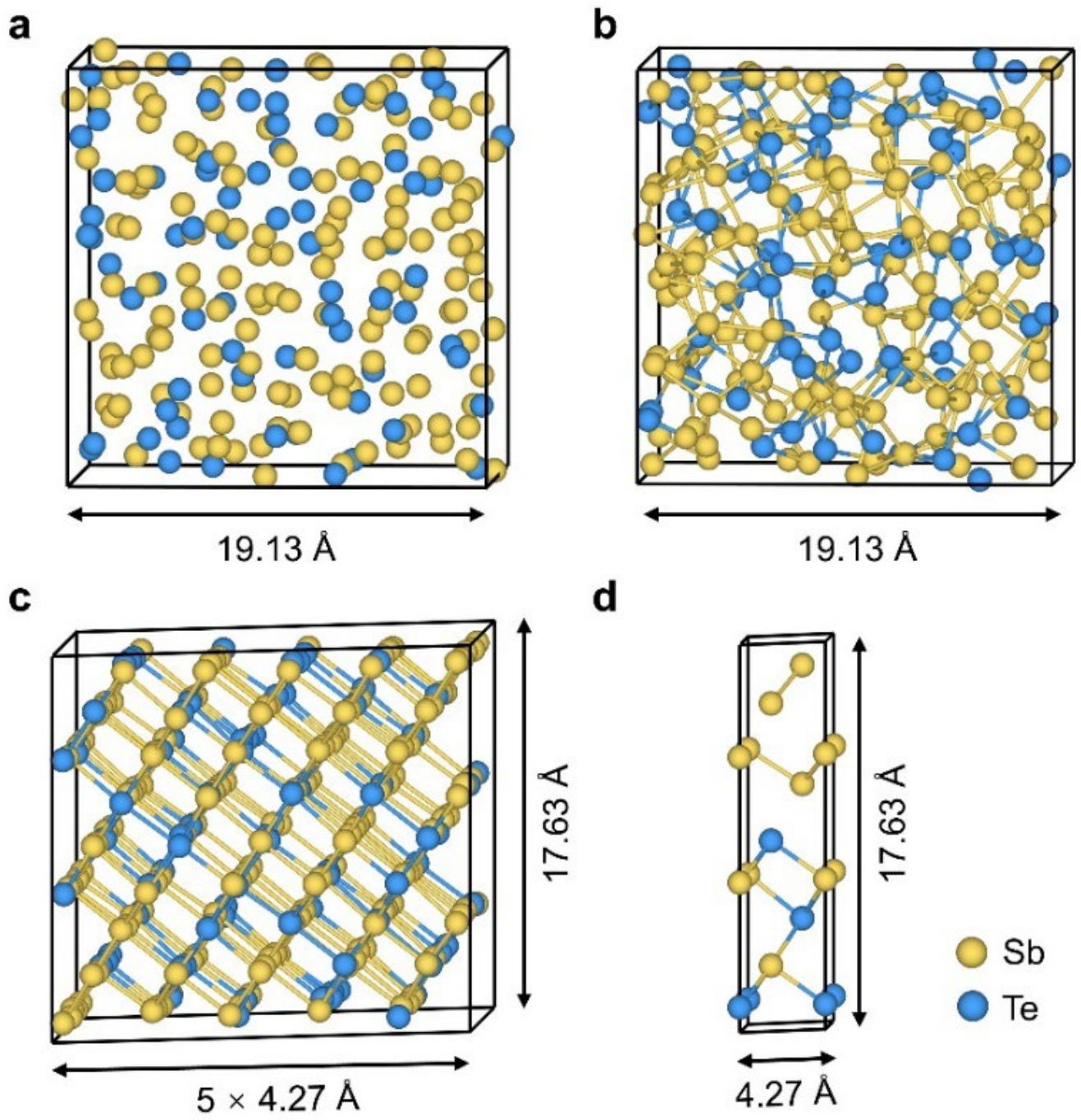

**Figure S1.** The full atomic structures of the four ST models shown in Figure 1a. The snapshot of AIMD-generated (a) liquid state at 1000 K (dynamical bonds are not visualized) and (b) amorphous state at 0 K. The DFT-optimized (c) metastable rhombohedral state and (d) ground state. The first two models are in cubic cell, and the latter two are in hexagonal cell.

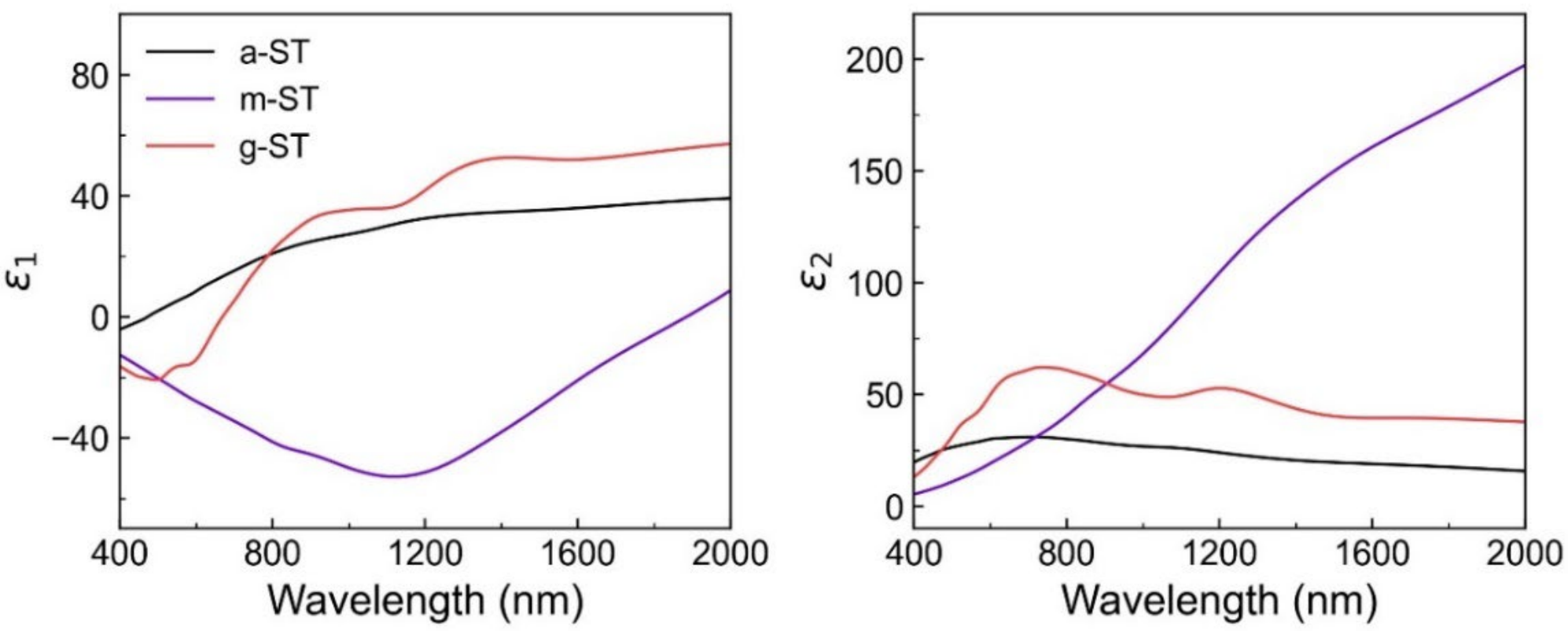

**Figure S2.** The DFT-calculated real part ($\varepsilon_1$) and imaginary part ($\varepsilon_2$) dielectric function of the amorphous model and the two crystalline models using HSE06 hybrid functional. The refractive index ($n$) and extinction coefficient ($k$) shown in Figure 1C were calculated via $n = \left(\frac{\sqrt{\varepsilon_1^2+\varepsilon_2^2}+\varepsilon_1}{2}\right)^{\frac{1}{2}}$ and $k = \left(\frac{\sqrt{\varepsilon_1^2+\varepsilon_2^2}-\varepsilon_1}{2}\right)^{\frac{1}{2}}$.

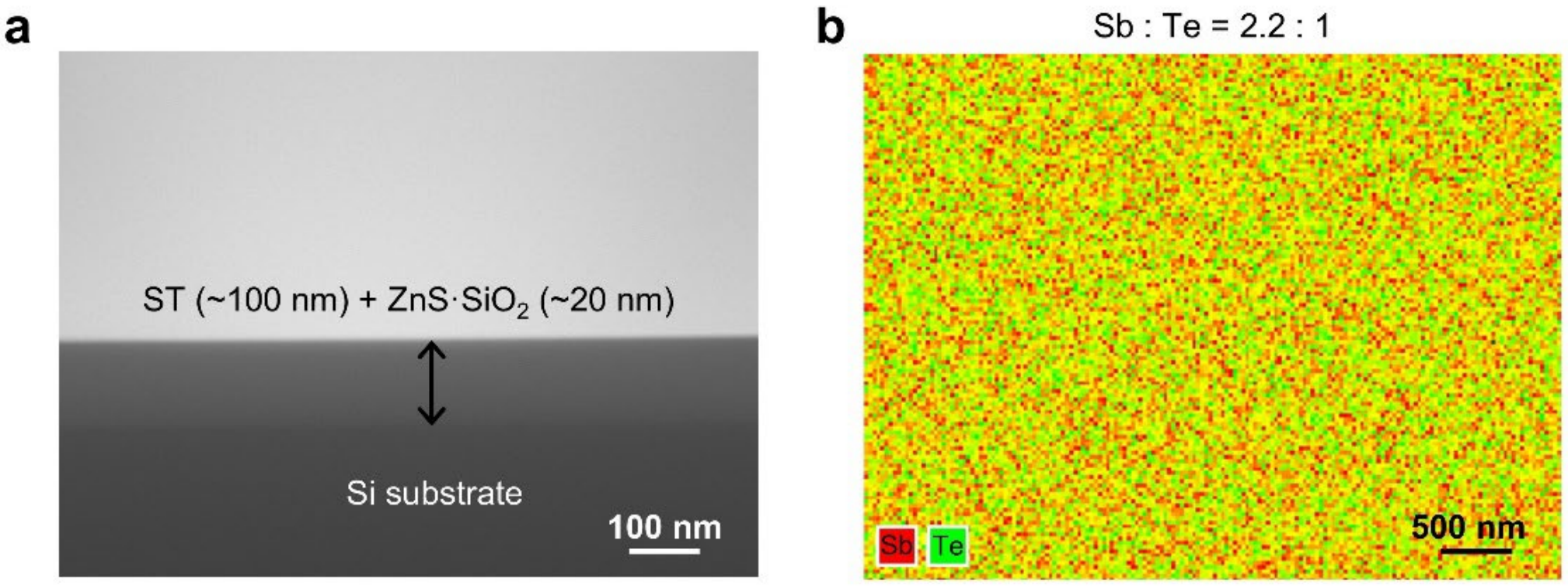

**Figure S3.** (a) The scanning electron microscopy (SEM) photograph for the cross-section of an as-deposited ST film. (b) The EDS map for an as-deposited film (in the top view).

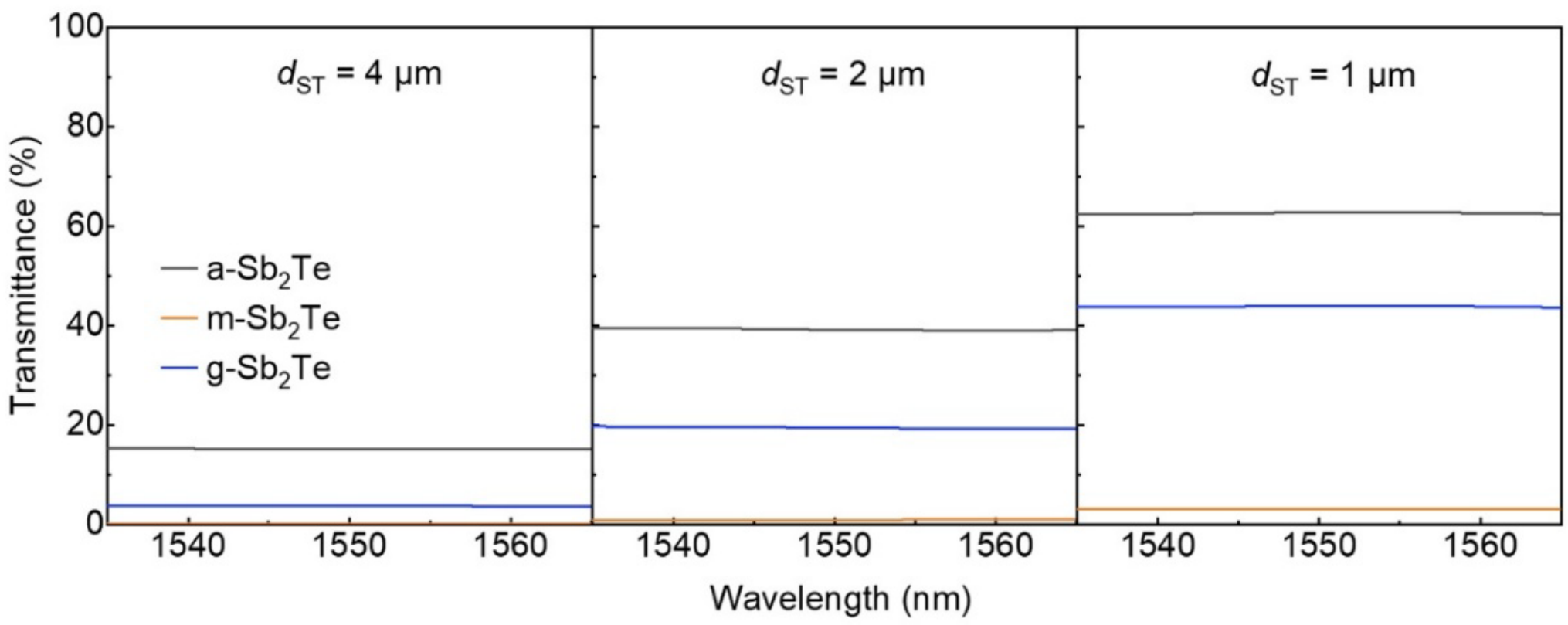

**Figure S4.** The FDTD simulations of ST-based waveguide devices using DFT-calculated *n* and *k* data. The difference between the g- and m-phases is bigger, because they represent the fully ordered and fully disordered crystalline states, respectively. The FDTD simulations shown in Figure 4 used the experimental *n* and *k* data of two intermediate crystalline states.

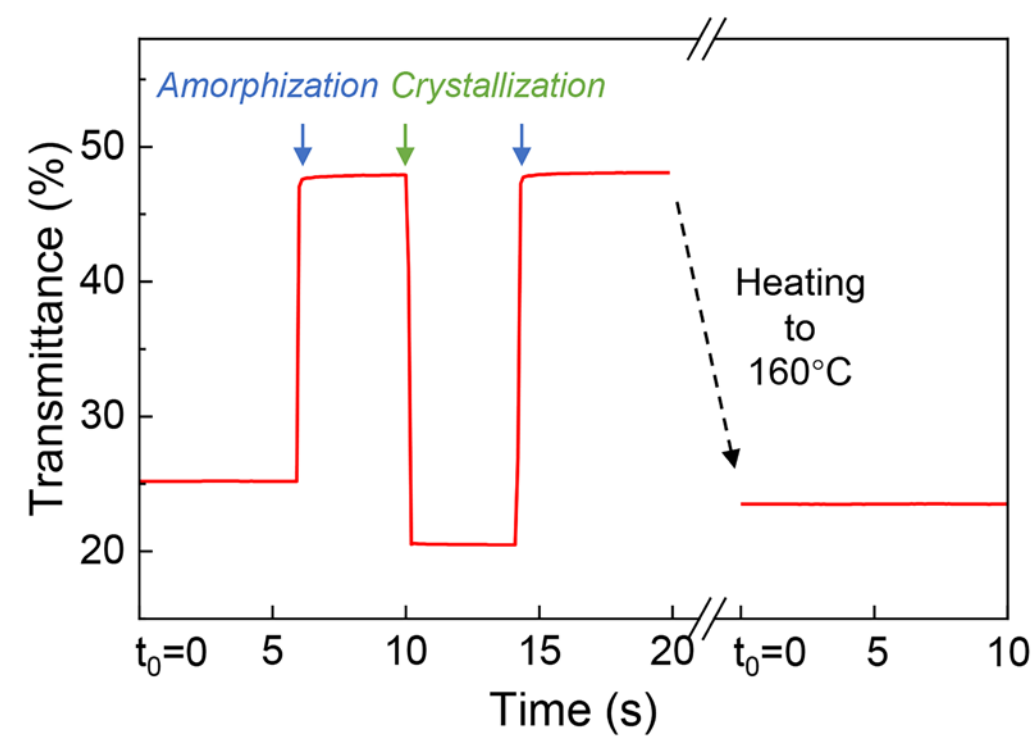

**Figure S5.** Waveguide measurement of the 1-μm device. After the second amorphization pulse, the ST device was heated to 160 °C and then cooled down to room temperature for optical measurement. This annealed state (as indicated by the black dashed arrow) shows a higher transmittance than the crystallized state obtained by optical pulsing (between the time interval of 10–15 s).

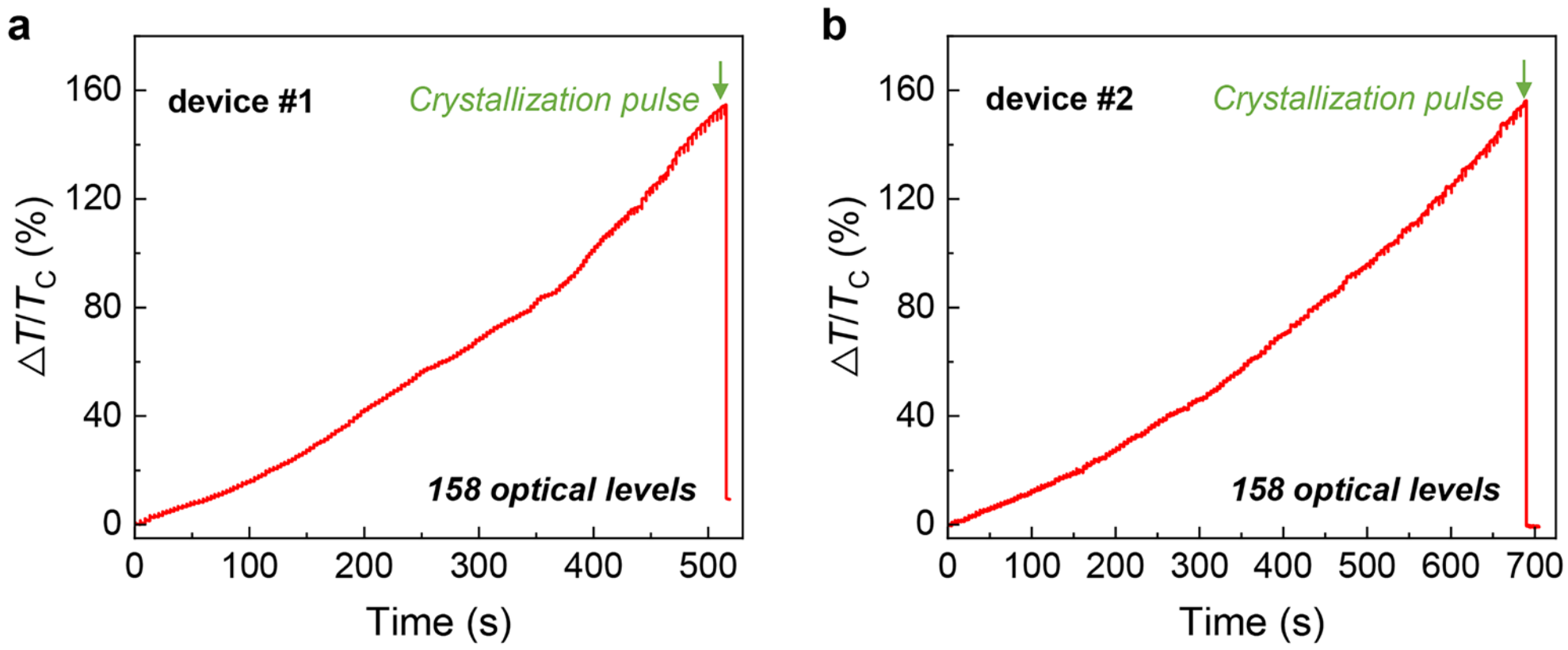

**Figure S6.** (a) The multilevel operation shown in the main text. (b) Another device with the same length and thickness of $Sb_2Te$ thin film ($d_{ST}$ = 1 μm and $h_{ST}$ = 10 nm) was fabricated and programmed separately. The multiple programming ability was well reproduced.

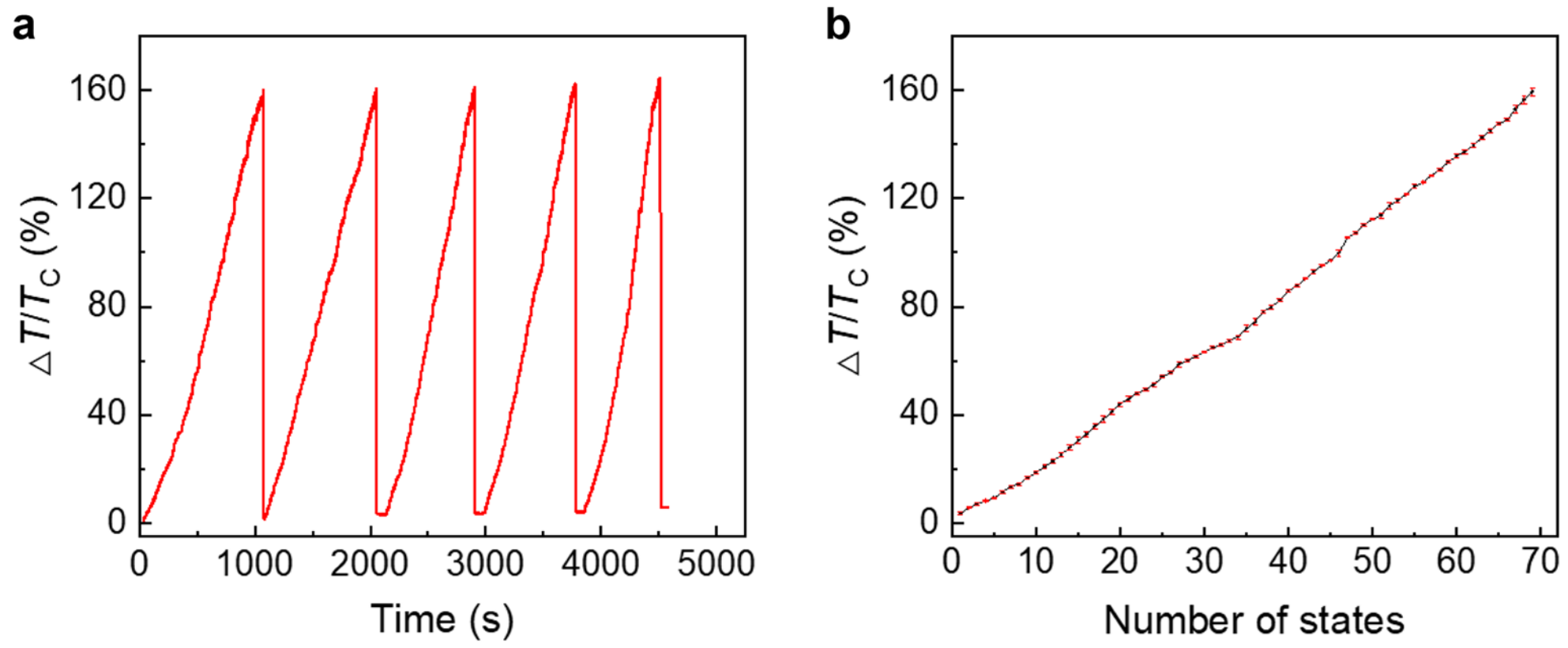

**Figure S7**. (a) Repeated multilevel operation of one ST-based waveguide device. After the partial amorphization operation, a long laser pulse was sent to crystallize the device. Such operation was repeated for 5 times. (b) The average transmittance values with standard deviation for the five programming cycles.

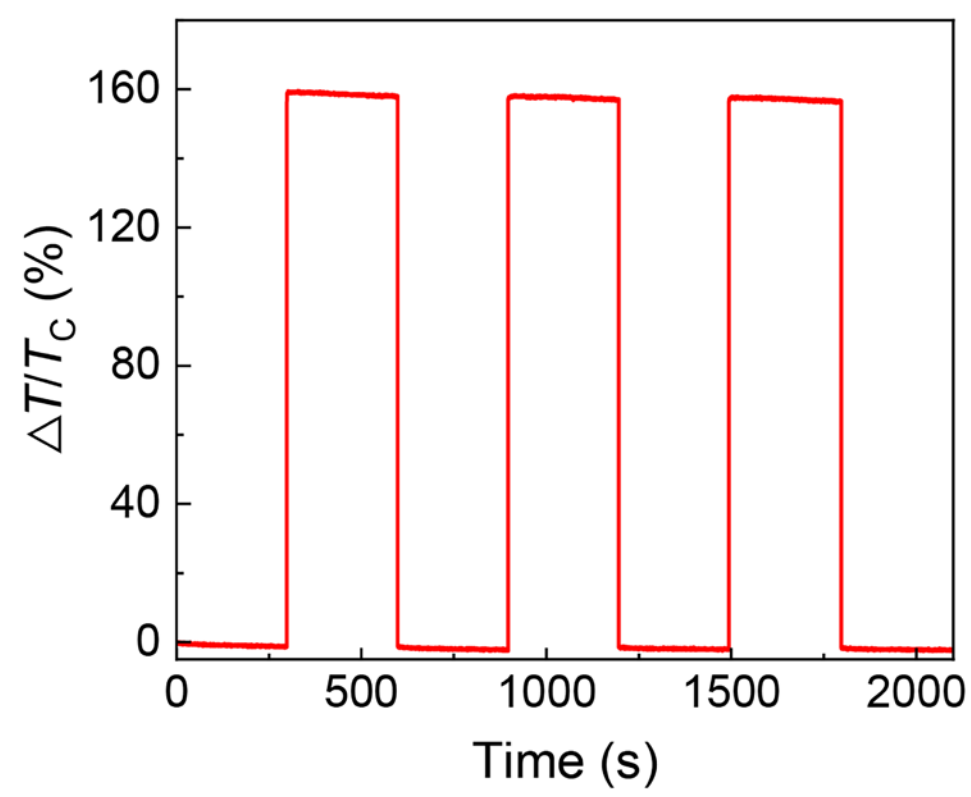

**Figure S8.** The high- and low-transmissive levels of the ST-based waveguide device are robust over time.

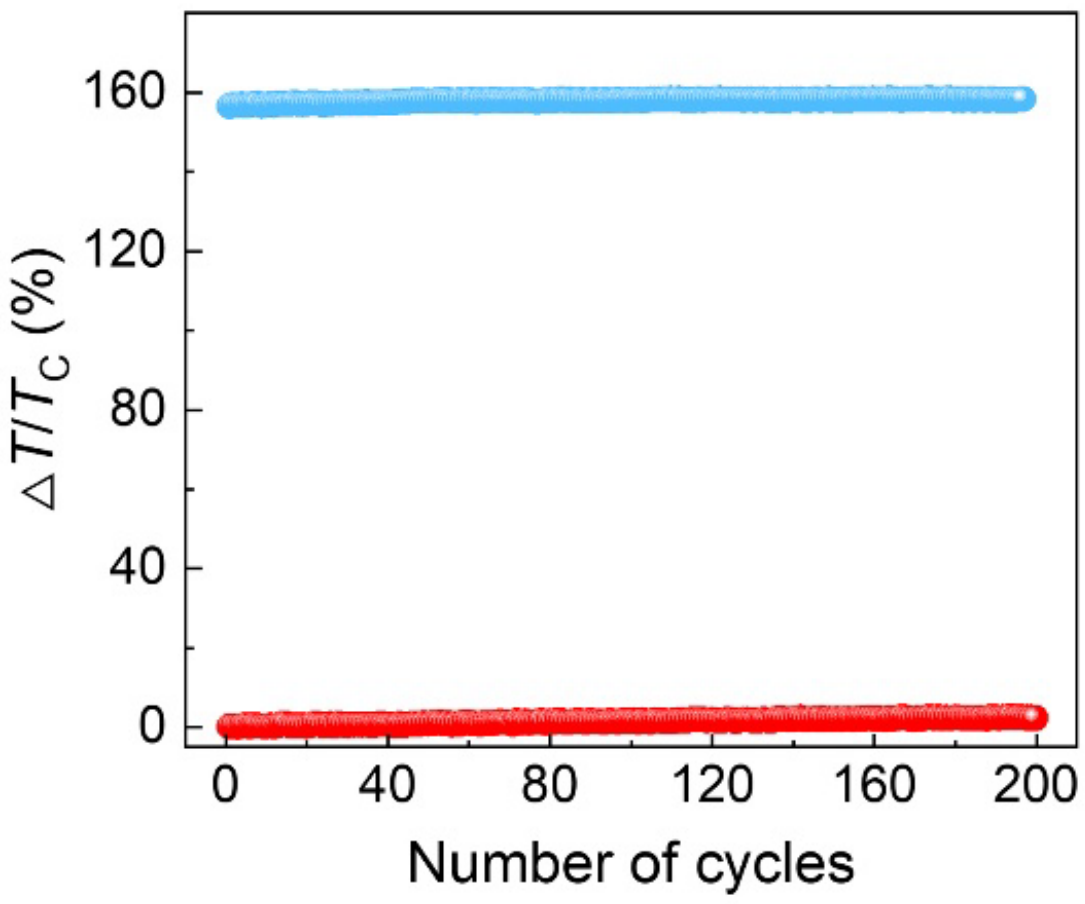

**Figure S9.** Repetitive optical switching with large optical window over write/erase 200 cycles.

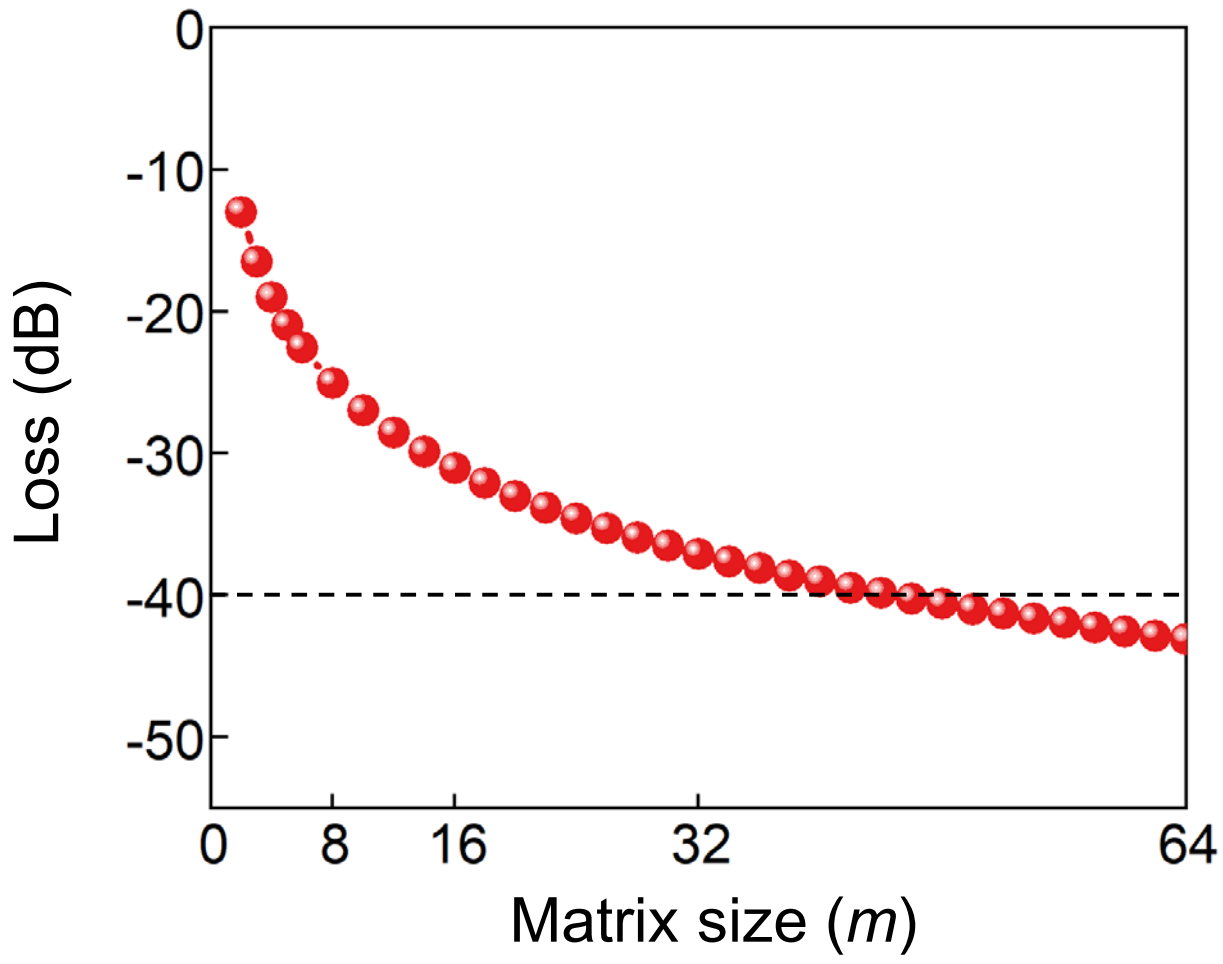

**Figure S10.** Estimation of the scalability of $Sb_2Te$ in a potential crossbar array. The loss of the crossbar array expressed as $D+10\times\log_{10}(1/m^2)$ is contributed by the insertion loss ($D$) of the PCM devices (−7.0 dB for a 1-μm-long $Sb_2Te$ device) and the light splitting ratio ($1/m$) of the crossbar array (Ref. 60). When the loss reaches −40 dB, the estimated array size is 45×45.

|  | Energy consumption | Write/Erase pulse widths | Optical contrast | Number of optical levels |
| --- | --- | --- | --- | --- |
| $Ge_2Sb_2Te_5$ Ref 13 | 430 pJ | 100/700 ns | 21% | 8 |
| $Ge_2Sb_2Te_5$ Ref 14 | 680 pJ | 50/250 ns | 28% | 34 |
| $Ge_2Sb_2Te_5$ Ref 15 | 917 pJ | 25/125 ns | 29% | 13 |
| $Ge_2Sb_2Te_5$ Ref 18 | N.A. | 10/100 ns | 183% | 65 |
| $Ag_3In_4Sb_{76}Te_{17}$ Ref 47 | 1.5 nJ | 20/70 ns | 20% | 45 |
| This work | 1072 pJ | 100/1025 ns | 156% | 158 |

**Table S1**. Benchmarks on performances of standard all-optical waveguide memory devices, without considering additional light nor electrical pump sources.